# Construction of Frequency Hopping Sequence Set Based upon Generalized Cyclotomy

Fang Liu, Daiyuan Peng, Zhengchun Zhou, and Xiaohu Tang

***Abstract:*** Frequency hopping (FH) sequences play a key role in frequency hopping spread spectrum communication systems. It is important to find FH sequences which have simultaneously good Hamming correlation，large family size and large period. In this paper, a new set of FH sequences with large period is proposed, and the Hamming correlation distribution of the new set is investigated. The construction of new FH sequences is based upon Whiteman's generalized cyclotomy. It is shown that the proposed FH sequence set is optimal with respect to the average Hamming correlation bound.

## 1   Introduction

Frequency hopping code division multiple access (FH-CDMA) is widely used in modern communication systems such as Bluetooth, ultra-wideband (UWB), military or radar applications, etc. In FH-CDMA systems, the receiver is confronted with the interference caused by undesired signals when it attempts to demodulate one of the signals from several transmitters. Generally, it is very desirable to keep the mutual interference, or the Hamming crosscorrelations and the out-of-phase Hamming autocorrelations of the frequency hopping (FH) sequences employed, as low as possible. On the other hand, it is also preferred to have more FH sequences accommodating more distinct users. The processing gain [1][2] of an FH system is clearly lost if the jammer knows the FH sequences being used. To prevent the jammer from storing and replaying to his advantage the FH sequences currently employed, it is necessary that the period $L$ of the FH sequences be large. As a consequence, the need for finding FH sequences which have simultaneously low Hamming correlation, large family size and large period is therefore well motivated.

The authors are with the Key Laboratory of Information Coding and Transmission, Southwest Jiaotong University, Chengdu, Sichuan, People's Republic of China. E-mail: hmimy5416@163.com

There are two kinds of measurement for the Hamming correlation of FH sequences: one is the maximum Hamming correlation [3][4] and another is the average Hamming correlation [5][6]. In recent years, the designs of FH sequences remain of great interest. Among of them, most have been devoted to the maximum Hamming correlation property [7][8][9][10][11]. While the average Hamming correlation indicates the average error (or interference) performance of the FH-CDMA systems, the design of optimal FH sequences with respect to the optimal average Hamming correlation property is very meaningful as well. However, up to now, only a few results on the average Hamming correlation of FH sequences have been reported [5][6][12][13].

A generalized cyclotomy with respect to $n = pq$ was introduced by Whiteman [14], where $p$ and $q$ are two different odd primes. When $n$ is a prime, it is referred to as classical cyclotomy. Some optimal or near-optimal FH sequence sets with respect to the maximum Hamming correlation bound were constructed based on classical cyclotomy [7][9][15][16]. Whiteman's generalized cyclotomy has been widely applied to design difference sets [14][17], as well to construct binary sequences with good correlation properties (but, not Hamming correlation) [18][19][20]. In this paper, we construct a new set of FH sequences based on Whiteman's generalized cyclotomy and investigate the average Hamming correlation of the FH sequence set. It is shown that the set of FH sequences is an optimal average Hamming correlation set.

The outline of this paper is as follows. In Section 2, we give some preliminaries on FH sequences, and review some bounds on the maximum and average Hamming correlation, respectively. In Section 3, we introduce the definition and some fundamental properties of Whiteman's generalized cyclotomy and the corresponding cyclotomic number. In Section 4, we give some basic lemmas that are needed to prove our main results. In Section 5, we focus on a new construction of the FH sequence set, and determine the Hamming correlation value of the FH sequences. Finally, we conclude this paper in Section 6.

## 2  Preliminaries

Let $F=\{f_0, f_1, \ldots, f_{v-1}\}$ be a set of available frequencies called a frequency library. Let $U$ be a set of $M$ FH sequences of length $L$ over $F$. Given two sequences $X=\{x_0, x_1, \ldots, x_{L-1}\}$ and $Y=\{y_0, y_1, \ldots, y_{L-1}\}$ in $U$, the periodic Hamming crosscorrelation function of $X$ and $Y$ is defined

by

$$H_{X,Y}(\tau) = \sum_{t=0}^{L-1} h[x_t, y_{t+\tau}], \ 0 \leq \tau \leq L-1$$

where $h[x_t, y_{t+\tau}]=1$ if $x_t = y_{t+\tau}$, and 0 otherwise, and the subscript addition is calculated modulo $L$. When $X=Y$, $H_{X,Y}(\tau)$ is called the Hamming autocorrelation function of $X$. In this case, we denote by $H_X(\tau)$.

The maximum Hamming autocorrelation sidelobe $H(X)$ of $X$ and the maximum Hamming crosscorrelation $H(X, Y)$ between $X$ and $Y$ are defined, respectively, by

$$H(X) = \max_{1 \leq \tau < L} \{H_X(\tau)\},$$
$$H(X,Y) = \max_{0 \leq \tau < L} \{H_{X,Y}(\tau)\}.$$

To evaluate single FH sequence, Lempel and Greenberger established the first bound in 1974, known as the Lempel-Greenberger bound.

*Lemma 1(The Lempel-Greenberger bound [3]):* For any FH sequence $X$ of length $L$ over $F$ with $|F|=v$,

$$H(X) \geq \left\lceil \frac{(L-b)(L+b-v)}{v(L-1)} \right\rceil \tag{1}$$

where $b$ denotes the nonnegative residue of $L$ modulo $v$, and $\lceil x \rceil$ denotes the smallest integer greater than or equal to $x$.

For any given FH sequence set $U$, the maximum Hamming autocorrelation sidelobe $H_a(U)$ and the maximum Hamming crosscorrelation $H_c(U)$ are defined, respectively, by

$$H_a(U) = \max_{X \in U} \{H(X)\},$$
$$H_c(U) = \max_{X,Y \in U, X \neq Y} \{H(X,Y)\}.$$

In 2004, Peng and Fan took account of the number of sequences in the family and then developed the following bound.

*Lemma 2 ([4]):* Let $U$ be a set of $M$ FH sequences of length $L$ over a frequency slot set $F$ with $|F|=v$, and $I = \lfloor LM/v \rfloor$, where $\lfloor x \rfloor$ denotes the largest integer less than or equal to $x$. Then

$$(L-1)vH_a(U) + (M-1)LvH_c(U) \geq (LM-v)L. \tag{2}$$

Another important performance indicator of the FH sequences is the average Hamming correlation defined by

*Definition 1 ([6]):* Let $U$ be a set of $M$ FH sequences of length $L$ over a given frequency slot

set *F* with size *v*, we call

$$S_a(U) = \sum_{X \in U,\ 1 \le \tau \le L-1} H_X(\tau), \qquad (3)$$

$$S_c(U) = \frac{1}{2} \sum_{X,Y \in U, X \ne Y, 0 \le \tau \le L-1} H_{X,Y}(\tau) \qquad (4)$$

as the overall number of Hamming autocorrelation and Hamming crosscorrelation of *U* respectively, and call

$$A_a(U) = \frac{S_a(U)}{M(L-1)}, \qquad (5)$$

$$A_c(U) = \frac{2S_c(U)}{LM(M-1)} \qquad (6)$$

as the average Hamming autocorrelation and the average Hamming crosscorrelation of *U* respectively.

For simplicity, we denote $H_a = H_a(U)$, $H_c = H_c(U)$, $S_a = S_a(U)$, $S_c = S_c(U)$, $A_a = A_a(U)$ and $A_c = A_c(U)$.

In 2008, Peng *et al.* derived the following theoretical limit which gave a bounded relation among the parameters $v$, $L$, $M$, $A_a$ and $A_c$.

*Lemma 3 ([6]):* Let *U* be a set of *M* FH sequences of length *L* over a given frequency slot set *F* with size *v*. Let $A_a$ and $A_c$ be the average Hamming autocorrelation and the average Hamming crosscorrelation of *U*, respectively. Then

$$\frac{A_a}{L(M-1)} + \frac{A_c}{(L-1)} \ge \frac{LM - v}{v(L-1)(M-1)}. \qquad (7)$$

Hereafter, we use the following definitions.

1) An FH sequence $X \in U$ is called optimal if the Lempel-Greenberger bound in Lemma 1 is met.

2) An FH sequence set *U* is an optimal set with respect to the maximum Hamming correlation bound if $H_a$ and $H_c$ of *U* is a pair of the minimum integer solutions of inequality (2).

3) An FH sequence set *U* is an optimal set with respect to the average Hamming correlation bound if the parameters $v$, $L$, $M$, $A_a$, and $A_c$ of *U* satisfy inequality (7) with equality.

## 3  Generalized cyclotomy and cyclotomic number

Let $p$ and $q$ be two distinct primes with $\gcd(p-1, q-1)=e$. According to the Chinese Remainder Theorem, there exists a common primitive root of $p$ and $q$, say $g$. Let $x$ be an integer satisfying the simultaneous congruences

$$\begin{cases} x \equiv g \pmod{p}, \\ x \equiv 1 \pmod{q}. \end{cases} \qquad (8)$$

Let $d=(p-1)(q-1)/e$, $f_1=(p-1)/e$, $f_2=(q-1)/e$ and $L=pq$. Thus, we can get a multiplicative subgroup of the residue class ring $\mathbf{Z}_L$, as follows [14]

$$\mathbf{Z}_L^* = \{g^s x^i : s = 0,1,\ldots,d-1;\ i = 0,1,\ldots,e-1\}$$

where $\mathbf{Z}_L^*$ denotes the set of all invertible elements of $\mathbf{Z}_L$.

Whiteman's generalized cyclotomic classes $D_i$, $0 \le i \le e-1$, of order $e$ are defined by

$$D_i = \{g^s x^i : s=0,1,\ldots,d-1\}$$

where the multiplication is performed modulo $L$. Obviously, $\mathbf{Z}_L^* = \bigcup_{i=0}^{e-1} D_i$.

Define

$$\begin{aligned} P &= \{p, 2p, \ldots, (q-1)p\}, \\ Q &= \{q, 2q, \ldots, (p-1)q\}, \\ R &= \{0\}. \end{aligned}$$

Let $H$ be a subset of $\mathbf{Z}_L$ and $a$ be an element of $\mathbf{Z}_L$. Define

$$H + a = \{h + a : h \in H\}, \qquad a \cdot H = \{a \cdot h : h \in H\}.$$

For fixed $i$ and $j$ with $0 \le i, j \le e-1$, the corresponding generalized cyclotomic number of order $e$ is defined by

$$(i,j) = |(D_i + 1) \cap D_j|. \qquad (9)$$

Now we give two fundamental properties of the generalized cyclotomic number.

*Lemma 4 ([17]):* The generalized cyclotomic number defined in (9) has the following properties:

1)  $\qquad\qquad\qquad (i,j) = (e-i, j-i);$

2) $$\sum_{i=0}^{e-1}(i,j) = \frac{(p-2)(q-2)-1}{e} + \varepsilon_j$$

where

$$\varepsilon_j = \begin{cases} 1, & \text{if } j = 0 \\ 0, & \text{otherwise.} \end{cases}$$

## 4 Basic Lemmas

In this section, we will give some useful lemmas for determining the Hamming correlation of our FH sequence set defined in the next section.

*Lemma 5 ([21]):*

$$\sum_{i=0}^{e-1}(k+i,i) = \begin{cases} \frac{(p-2)(q-2)-1}{e} + 1, & \text{if } k = 0 \\ \frac{(p-2)(q-2)-1}{e}, & \text{otherwise.} \end{cases} \quad (10)$$

*Lemma 6 ([21]):* For any $k \in \mathbf{Z}_e \setminus \{0\}$, we have

1) $$\sum_{i=0}^{e-1}|(D_i + w) \cap D_i| = \begin{cases} ef_1(f_2-1), & \text{if } w \in P \\ ef_2(f_1-1), & \text{if } w \in Q \\ \frac{(p-2)(q-2)-1}{e} + 1, & \text{if } w \in \mathbf{Z}_L^*; \end{cases}$$

2) $$\sum_{i=0}^{e-1}|(D_{i+k} + w) \cap D_i| = \begin{cases} ef_1 f_2, & \text{if } w \in P \cup Q \\ \frac{(p-2)(q-2)-1}{e}, & \text{if } w \in \mathbf{Z}_L^*. \end{cases}$$

*Lemma 7:*

$$|((Q \cup R) + w) \cap (Q \cup R)| = \begin{cases} 0, & \text{if } w \in P \cup \mathbf{Z}_L^* \\ p, & \text{if } w \in Q \cup R \end{cases}$$

and

$$|(P + w) \cap P| = \begin{cases} q-2, & \text{if } w \in P \\ q-1, & \text{if } w = 0 \\ 0, & \text{otherwise.} \end{cases}$$

*Proof:* This lemma is obvious, so we omit the proof. ∎

*Lemma 8:* Given $0 \leq i \leq e-1$, then

$$|(D_i+w)\cap(Q\cup R)|=\begin{cases} 0, & \text{if } w\in Q\cup R \\ f_1, & \text{if } w\in P\cup \mathbf{Z}_L^* \end{cases}$$

and

$$|(D_i+w)\cap(P\cup R)|=\begin{cases} 0, & \text{if } w\in P\cup R \\ f_2, & \text{if } w\in Q\cup \mathbf{Z}_L^*. \end{cases}$$

*Proof:* We only prove the first equation since the second one is similar.

When $w\in Q\cup R$, $|(D_i+w)\cap(Q\cup R)|=0$ is clear. As for $w\in P\cup \mathbf{Z}_L^*$, an element $z=g^s x^i+w\in(Q\cup R)$, $0\leq s\leq d-1$, $0\leq i\leq e-1$ if and only if

$$g^s+w\equiv 0(\bmod\ q) \tag{11}$$

in which we make use of the fact that $x\equiv 1(\bmod\ q)$. Let $s=t(q-1)+s_1$, where $0\leq t\leq f_1-1$, $0\leq s_1\leq q-1$. Obviously, only one $s_1$ in $\mathbf{Z}_q$ satisfies (11). Then, there are $f_1$ solutions $0\leq s\leq d-1$ to (11). Therefore, the number of solutions of (11) with $0\leq s\leq d-1$ is $f_1$. ∎

*Lemma 9 ([21]):* $-1\in D_0$ if $|f_1-f_2|$ is even, and $-1\in D_{e/2}$ if $|f_1-f_2|$ is odd.

*Lemma 10:* For any $0\leq i\leq e-1$, we have

1) 
$$|(D_i+w)\cap P|=\begin{cases} 0, & \text{if } w\in P\cup R \\ f_2, & \text{if } w\in Q \\ f_2-1, & \text{if } w\in D_i \text{ and } |f_1-f_2| \text{ is even} \\ f_2, & \text{if } w\in \mathbf{Z}_L^*\setminus D_i \text{ and } |f_1-f_2| \text{ is even} \\ f_2-1, & \text{if } w\in D_{e/2+i} \text{ and } |f_1-f_2| \text{ is odd} \\ f_2, & \text{if } w\in \mathbf{Z}_L^*\setminus D_{e/2+i} \text{ and } |f_1-f_2| \text{ is odd}; \end{cases} \tag{12}$$

2) 
$$|(P+w)\cap D_i|=\begin{cases} 0, & \text{if } w\in P\cup R \\ f_2, & \text{if } w\in Q\cup(\mathbf{Z}_L^*\setminus D_i) \\ f_2-1, & \text{if } w\in D_i. \end{cases} \tag{13}$$

*Proof:* For 1), note that

$$|(D_i+w)\cap P|=|(D_i+w)\cap(P\cup R)|-|(D_i+w)\cap R|.$$

By Lemma 9, we have

$$|(D_i+w)\cap R|=\begin{cases} 0, & \text{if } w\in P\cup Q\cup R \\ 1, & \text{if } w\in D_i \text{ and } |f_1-f_2| \text{ is even} \\ 0, & \text{if } w\in \mathbf{Z}_L^*\setminus D_i \text{ and } |f_1-f_2| \text{ is even} \\ 1, & \text{if } w\in D_{e/2+i} \text{ and } |f_1-f_2| \text{ is odd} \\ 0, & \text{if } w\in \mathbf{Z}_L^*\setminus D_{e/2+i} \text{ and } |f_1-f_2| \text{ is odd}. \end{cases} \tag{14}$$

Then the conclusion follows from Lemma 8 and Equation (14).

For 2), we have

$$|(P+w)\cap D_i|=|P\cap(D_i-w)|=|(P\cup R)\cap(D_i-w)|-|R\cap(D_i-w)|.$$

Applying Lemma 8, we arrive at the conclusion. ∎

*Lemma 11 ([18]):*

$$|(P+w)\cap(Q\cup R)|=\begin{cases}0,&\text{if }w\in Q\\1,&\text{if }w\in P\cup\mathbf{Z}_L^*.\end{cases}$$

## 5 New construction of FH sequences based on Whiteman's generalized cyclotomy

In this section, we construct a new set of FH sequences with optimal average Hamming correlation property.

Let

$$C_0 = D_0 \cup Q \cup R,$$
$$C_i = D_i, \text{ for } 1\leq i\leq e-1, i\neq e/2,$$
$$C_{e/2} = D_{e/2} \cup P.$$

Then, $\bigcup_{i=0}^{e-1} C_i = \mathbf{Z}_L$ and $C_i \cap C_j = \varnothing$ for $i\neq j$.

Let $X = \{x_0, x_1, \ldots, x_{L-1}\}$ be a sequence of length $L$ over a frequency slot set $F$. Then $\text{supp}_X(k)=\{t \mid x_t=k, 0\leq t\leq L-1\}$ is called the support of $k\in F$ in the sequence $X$.

*Definition 2:* Define an FH sequence set $U = \{X^{(i)}: i = 0,1,\ldots,e-1\}$ of length $L=pq$, where $X^{(i)} = \{x_0^{(i)}, x_1^{(i)}, \ldots, x_{L-1}^{(i)}\}$ is defined by

$$\text{supp}_{X^{(i)}}(j) = C_{j+i}, 0\leq j\leq e-1 \tag{15}$$

where $j+i$ is reduced modulo $e$.

Based on the lemmas in last section, we are now ready to determine the Hamming correlation properties of the FH sequence set $U$.

*Theorem 1:* Let $p$ and $q$ be different odd primes with $\gcd(p-1, q-1)=e$. Define $p=ef_1+1$ and $q=ef_2+1$, then the FH sequence set $U$ over $F$ have the following properties.

1) The family size $M=e$, the sequence length $L=pq$, and $|F|=e$.

2) The Hamming autocorrelation function of $X^{(k)}\in U$ for $0\leq k\leq e-1$ is given by

$$H_{X^{(k)}}(w) = \begin{cases} \dfrac{pq-1}{e} + \dfrac{p-q}{e} + q - p - 1, & \text{if } w \in P \\ \dfrac{pq-1}{e} + \dfrac{q-p}{e} + p - q + 1, & \text{if } w \in Q \\ \dfrac{pq-1}{e} - 1, & \text{if } w \in D_{e/2} \text{ and } |f_1 - f_2| \text{ is even} \\ \dfrac{pq-1}{e}, & \text{if } w \in D_0 \cup D_{e/2} \text{ and } |f_1 - f_2| \text{ is odd} \\ \dfrac{pq-1}{e} + 1, & \text{if } w \in D_0 \text{ and } |f_1 - f_2| \text{ is even} \\ \dfrac{pq-1}{e} + 1, & \text{if } w \in D_i \text{ for } i \ne 0, e/2. \end{cases}$$

3) The Hamming crosscorrelation function of any two distinct FH sequences $X^{(k)}$, $X^{(l)} \in U$ for $k \ne l$ is given by

3.1) When $l - k \equiv e/2 \pmod{e}$

$$H_{X^{(k)}, X^{(l)}}(w) = \begin{cases} 0, & \text{if } w = 0 \\ \dfrac{pq-1}{e} + \dfrac{p-q}{e} + 2, & \text{if } w \in P \\ \dfrac{pq-1}{e} + \dfrac{q-p}{e}, & \text{if } w \in Q \\ \dfrac{pq-1}{e} + 2, & \text{if } w \in D_{e/2} \text{ and } |f_1 - f_2| \text{ is even} \\ \dfrac{pq-1}{e} + 1, & \text{if } w \in D_0 \cup D_{e/2} \text{ and } |f_1 - f_2| \text{ is odd} \\ \dfrac{pq-1}{e}, & \text{if } w \in D_0 \text{ and } |f_1 - f_2| \text{ is even} \\ \dfrac{pq-1}{e} + 2, & \text{if } w \in D_i \text{ for } i \ne 0, e/2. \end{cases} \quad (16)$$

3.2) When $2(l-k) \equiv e/2 \pmod{e}$,

$$H_{X^{(k)}, X^{(l)}}(w) = \begin{cases} 0, & \text{if } w = 0 \\ \dfrac{pq-1}{e} + \dfrac{p-q}{e}, & \text{if } w \in P \\ \dfrac{pq-1}{e} + \dfrac{q-p}{e}, & \text{if } w \in Q \\ \dfrac{pq-1}{e} - 1, & \text{if } w \in D_{l-k} \cup D_{l-k+e/2} \text{ and } |f_1 - f_2| \text{ is even} \\ \dfrac{pq-1}{e} - 2, & \text{if } w \in D_{l-k} \text{ and } |f_1 - f_2| \text{ is odd} \\ \dfrac{pq-1}{e}, & \text{if } w \in D_{l-k+e/2} \text{ and } |f_1 - f_2| \text{ is odd} \\ \dfrac{pq-1}{e}, & \text{if } w \in D_i \text{ for } i \ne l-k, l-k+e/2. \end{cases} \quad (17)$$

3.3) When $2(l-k) \not\equiv e/2 \pmod{e}$ and $l - k \not\equiv e/2 \pmod{e}$,

$$H_{X^{(k)},X^{(l)}}(w) = \begin{cases} 0, & \text{if } w = 0 \\ \frac{pq-1}{e} + \frac{p-q}{e}, & \text{if } w \in P \\ \frac{pq-1}{e} + \frac{q-p}{e}, & \text{if } w \in Q \\ \frac{pq-1}{e}, & \text{if } w \in D_{l-k} \text{ and } |f_1 - f_2| \text{ is even} \\ \frac{pq-1}{e} - 1, & \text{if } w \in D_{l-k} \text{ and } |f_1 - f_2| \text{ is odd} \\ \frac{pq-1}{e} - 1, & \text{if } w \in D_{l-k+e/2} \text{ and } |f_1 - f_2| \text{ is even} \\ \frac{pq-1}{e}, & \text{if } w \in D_{l-k+e/2} \text{ and } |f_1 - f_2| \text{ is odd} \\ \frac{pq-1}{e}, & \text{if } w \in D_i \text{ for } i \neq l-k, l-k+e/2, k-l+e/2 \\ \frac{pq-1}{e} - 1, & \text{if } w \in D_{k-l+e/2}. \end{cases} \quad (18)$$

*Proof:* 1) is clear.

Concerning 2), the Hamming autocorrelation of $X^{(k)}$ at shift $w$ is

$$\begin{aligned} H_{X^{(k)}}(w) = & \sum_{i=0}^{e-1} |(D_i + w) \cap D_i| + |(D_0 + w) \cap (Q \cup R)| \\ & + |((Q \cup R) + w) \cap D_0| + |(P + w) \cap P| \\ & + |(D_{e/2} + w) \cap P| + |(P + w) \cap D_{e/2}| \\ & + |((Q \cup R) + w) \cap (Q \cup R)|. \end{aligned} \quad (19)$$

Then by Lemma 6, 7, 8, and 10, the result follows.

Regarding 3), for any FH sequences $X^{(k)}$, $X^{(l)} \in U$ with $k \neq l$ and $0 \leq k, l \leq e-1$, their Hamming crosscorrelation function at shift $w$ is given by

$$H_{X^{(k)},X^{(l)}}(w) = \sum_{i=0}^{e-1} |(C_{i+l} + w) \cap C_{i+k}|.$$

When $l-k \equiv e/2 \pmod{e}$, we have

$$\begin{aligned} H_{X^{(k)},X^{(l)}}(w) = & \sum_{i=0}^{e-1} |(D_{i+e/2} + w) \cap D_i| + |((Q \cup R) + w) \cap D_{e/2}| \\ & + |((Q \cup R) + w) \cap P| + |(D_0 + w) \cap P| \\ & + |(D_{e/2} + w) \cap (Q \cup R)| + |(P + w) \cap (Q \cup R)| \\ & + |(P + w) \cap D_0|. \end{aligned}$$

Applied Lemmas 6, 8, 10 and 11 to the above equation, the conclusion in 3.1) then follows.

While for any FH sequences $X^{(k)}$, $X^{(l)} \in U$ with $l-k \not\equiv e/2 \pmod{e}$, their Hamming crosscorrelation function at shift $w$ is given by

$$H_{X^{(k)}, X^{(l)}}(w) = \sum_{i=0}^{e-1} |(D_{i+l-k} + w) \cap D_i| + |(D_{l-k} + w) \cap (Q \cup R)|$$
$$+ |(D_{l-k+e/2} + w) \cap P| + |((Q \cup R) + w) \cap D_{k-l}|$$
$$+ |(P + w) \cap D_{k-l+e/2}|.$$

When $2(l-k) \equiv e/2 \pmod{e}$, it is easily verified that $|(P+w) \cap D_{k-l+e/2}| = |(P+w) \cap D_{l-k}|$. Therefore, the equation in 3.2) follows immediately from Lemmas 6, 8, and 10. Similarly, when $2(l-k) \not\equiv e/2 \pmod{e}$, from Lemma 6, 8, and 10, the desired result in 3.3) follows, which completes the proof. ∎

*Theorem 2:* The average Hamming autocorrelation and average Hamming crosscorrelation of the FH sequence set $U$ are respectively as follows

$$A_a(U) = \frac{S_a(U)}{M(L-1)}$$
$$= \frac{(pq-1)^2 + e(q^2+p^2) + e(1-pq) - 2eq - (q-1)^2 - (p-1)^2}{e(pq-1)}, \quad (20)$$

$$A_c(U) = \frac{2S_c(U)}{LM(M-1)}$$
$$= \frac{(e-1)(pq-1)^2 + 2ep(q-1) - (e-1)(q-1)^2 - (e-1)(p-1)^2}{pqe(e-1)}. \quad (21)$$

The FH sequence set $U$ is optimal with respect to the average Hamming correlation bound.

*Proof:* When $|f_1 - f_2|$ is even, according to the definitions of $S_a$ and $S_c$, we have

$$S_a = \sum_{0 \leq i \leq e-1, \, 1 \leq \tau \leq L-1} H_{X^{(i)}}(\tau)$$
$$= e\left\{(q-1)\left(\frac{pq-1}{e} + \frac{p-q}{e} + q - p - 1\right) + (p-1)\left(\frac{pq-1}{e} + \frac{q-p}{e} + p - q + 1\right)\right.$$
$$\left. + d\left(\frac{pq-1}{e} - 1\right) + d\left(\frac{pq-1}{e} + 1\right) + (e-2)d\left(\frac{pq-1}{e} + 1\right)\right\}$$
$$= (pq-1)^2 + e(q^2+p^2) + e(1-pq) - 2eq - (q-1)^2 - (p-1)^2$$

and

$$2S_c = \sum_{\substack{0 \leq i, j \leq e-1, \\ 0 \leq \tau \leq L-1, i \neq j}} H_{X^{(i)}, X^{(j)}}(\tau)$$
$$= \sum_{\substack{0 \leq i, j \leq e-1, 0 \leq \tau \leq L-1, \\ i-j \equiv e/2 \pmod{e}}} H_{X^{(i)}, X^{(j)}}(\tau) + \sum_{\substack{0 \leq i, j \leq e-1, 0 \leq \tau \leq L-1, \\ 2(i-j) \equiv e/2 \pmod{e}}} H_{X^{(i)}, X^{(j)}}(\tau) + \sum_{\substack{0 \leq i, j \leq e-1, 0 \leq \tau \leq L-1, 2(i-j) \not\equiv e/2 \\ \pmod{e}, \, i-j \not\equiv e/2 \pmod{e}, i \neq j}} H_{X^{(i)}, X^{(j)}}(\tau).$$

From Theorem 1, then

$$2S_c = \sum_{\substack{0\le i,j\le e-1, 0\le \tau\le L-1,\\ i-j\equiv e/2\,(\text{mod }e)}} \left\{(q-1)\left(\frac{pq-1}{e}+\frac{p-q}{e}+2\right)+(p-1)\left(\frac{pq-1}{e}+\frac{q-p}{e}\right)\right.$$

$$\left.+d\left(\frac{pq-1}{e}+2\right)+d\frac{pq-1}{e}+(e-2)d\left(\frac{pq-1}{e}+2\right)\right\}$$

$$+\sum_{\substack{0\le i,j\le e-1, 0\le \tau\le L-1,\\ 2(i-j)\equiv e/2\,(\text{mod }e)}} \left\{(q-1)\left(\frac{pq-1}{e}+\frac{p-q}{e}\right)+(p-1)\left(\frac{pq-1}{e}+\frac{q-p}{e}\right)\right.$$

$$\left.+2d\left(\frac{pq-1}{e}-1\right)+(e-2)d\frac{pq-1}{e}\right\}$$

$$+\sum_{\substack{0\le i,j\le e-1, 0\le \tau\le L-1, 2(i-j)\ne e/2\,(\text{mod }e),\\ i-j\ne e/2\,(\text{mod }e), i\ne j}} \left\{(q-1)\left(\frac{pq-1}{e}+\frac{p-q}{e}\right)+(p-1)\left(\frac{pq-1}{e}+\frac{q-p}{e}\right)\right.$$

$$\left.+2d\left(\frac{pq-1}{e}-1\right)+(e-2)d\frac{pq-1}{e}\right\}$$

$$=(e-1)(pq-1)^2+2ep(q-1)-(e-1)(q-1)^2-(e-1)(p-1)^2.$$

Applying (5) and (6), we obtain (20) and (21).

Similarly, when $|f_1-f_2|$ is odd, we obtain the same average Hamming autocorrelation and average Hamming crosscorrelation. By applying (20) and (21) to (7), it follows that

$$\frac{A_a}{L(M-1)}+\frac{A_c}{(L-1)} = \frac{(pq-1)^2+e(q^2+p^2)+e(1-pq-2q)-(q-1)^2-(p-1)^2}{e(e-1)pq(pq-1)}$$

$$+\frac{(e-1)(pq-1)^2+2epq-2ep-(e-1)((q-1)^2+(p-1)^2)}{e(e-1)pq(pq-1)}$$

$$=\frac{1}{e-1}\ge\frac{LM-v}{v(L-1)(M-1)}=\frac{pqe-e}{e(pq-1)(e-1)}=\frac{1}{e-1}.$$

Thus, the FH sequence set $U$ is an optimal average Hamming correlation set. ∎

*Example 1:* Let $p=5$, $q=17$, then $e=4$, $d=16$, $f_1=1$, $f_2=4$ and $|f_1-f_2|=3$. The FH sequences of $U$ are

$X^{(0)}$ = {00102210302121120010202232202123110222112113323300203312302021 00222
302233032212320232};

$X^{(1)}$ = {11213321013232231121313303313230221333223220030011310023013 13211333
013300103323031303};

$X^{(2)}$ = {22320032120303302232020010020301332000330331101122021130120 20322000
120011210030102010};

$X^{(3)}$ = {33031103231010013303131121131012003111001002212233132201231 31033111
231122321101213121}.

The Hamming autocorrelation of $X^{(i)}$ for $i = 0, 1, 2, 3$ is

$H_{X^{(i)}}$ = {85, 21, 22, 21, 21, 29, 22, 21, 22, 21, 29, 22, 21, 22, 22, 29, 21, 13, 22, 21, 29, 21,
21, 21, 22, 29, 21, 21, 21, 22, 29, 22, 22, 22, 13, 29, 21, 21, 22, 22, 29, 22, 22, 22,
22, 29, 22, 22, 21, 21, 29, 13, 22, 22, 22, 29, 22, 21, 21, 21, 29, 22, 21, 21, 21, 29,
21, 22, 13, 21, 29, 22, 22, 21, 22, 29, 21, 22, 21, 22, 29, 21, 21, 22, 21}.

The Hamming crosscorrelation is

$H_{X^{(0)}, X^{(1)}}$ = {0, 21, 19, 21, 21, 18, 19, 21, 21, 21, 18, 19, 21, 19, 19, 18, 21, 24, 19, 21, 18, 21,
21, 21, 21, 18, 21, 21, 21, 19, 18, 21, 19, 19, 24, 18, 21, 21, 19, 19, 18, 19, 19, 21,
21, 18, 21, 21, 21, 21, 18, 24, 21, 21, 19, 18, 21, 21, 21, 21, 8, 19, 21, 21, 21, 18,
21, 21, 24, 21, 18, 21, 21, 21, 21, 18, 21, 19, 21, 21, 18, 21, 21, 21, 21};

$H_{X^{(0)}, X^{(2)}}$ = {0, 22, 23, 22, 22, 20, 23, 22, 23, 22, 20, 23, 22, 23, 23, 20, 22, 24, 23, 22, 20, 22, 22,
22, 23, 20, 22, 22, 22, 23, 20, 23, 23, 23, 24, 20, 22, 22, 23, 23, 20, 23, 23, 23, 23, 20,
23, 23, 22, 22, 20, 24, 23, 23, 23, 20, 23, 22, 22, 22, 20, 23, 22, 22, 22, 20, 22, 23, 24,
22, 20, 23, 23, 22, 23, 20, 22, 23, 22, 23, 20, 22, 22, 23, 22};

$H_{X^{(0)}, X^{(3)}}$ = {0, 21, 21, 21, 21, 18, 21, 21, 19, 21, 18, 21, 21, 21, 21, 18, 21, 24, 21, 21, 18, 21, 21,
21, 19, 18, 21, 21, 21, 21, 18, 19, 21, 21, 24, 18, 21, 21, 21, 21, 18, 21, 21, 19, 19, 18,
19, 19, 21, 21, 18, 24, 19, 19, 21, 18, 19, 21, 21, 21, 18, 21, 21, 21, 21, 18, 21, 19, 24,
21, 18, 19, 19, 21, 19, 18, 21, 21, 21, 19, 18, 21, 21, 19, 21};

$H_{X^{(1)}, X^{(2)}}$ = {0, 21, 19, 21, 21, 18, 19, 21, 21, 21, 18, 19, 21, 19, 19, 18, 21, 24, 19, 21, 18, 21, 21,
21, 21, 18, 21, 21, 21, 19, 18, 21, 19, 19, 24, 18, 21, 21, 19, 19, 18, 19, 19, 21, 21, 18,
21, 21, 21, 21, 18, 24, 21, 21, 19, 18, 21, 21, 21, 21, 18, 19, 21, 21, 21, 18, 21, 21, 24,
21, 18, 21, 21, 21, 21, 18, 21, 19, 21, 21, 18, 21, 21, 21, 21};

$H_{X^{(1)}, X^{(3)}}$ = {0, 22, 23, 22, 22, 20, 23, 22, 23, 22, 20, 23, 22, 23, 23, 20, 22, 24, 23, 22, 20, 22, 22,
22, 23, 20, 22, 22, 22, 23, 20, 23, 23, 23, 24, 20, 22, 22, 23, 23, 20, 23, 23, 23, 23, 20,
23, 23, 22, 22, 20, 24, 23, 23, 23, 20, 23, 22, 22, 22, 20, 23, 22, 22, 22, 20, 22, 23, 24,
22, 20, 23, 23, 22, 23, 20, 22, 23, 22, 23, 20, 22, 22, 23, 22};

$H_{X^{(2)}, X^{(3)}}$ = {0, 21, 19, 21, 21, 18, 19, 21, 21, 21, 18, 19, 21, 19, 19, 18, 21, 24, 19, 21, 18, 21, 21,
21, 21, 18, 21, 21, 21, 19, 18, 21, 19, 19, 24, 18, 21, 21, 19, 19, 18, 19, 19, 21, 21, 18,
21, 21, 21, 21, 18, 24, 21, 21, 19, 18, 21, 21, 21, 21, 18, 19, 21, 21, 21, 18, 21, 21, 24,
21, 18, 21, 21, 21, 21, 18, 21, 19, 21, 21, 18, 21, 21, 21, 21}.

The average Hamming auto- and cross-correlation are $473/21$ and $5248/255$ respectively.

The sequence set $U$ is optimal with respect to the average Hamming correlation bound.

## 6  Conclusion

The average Hamming correlation is an important performance indicator of the FH sequences.

A new FH sequence set with length *pq*, family size *e* and frequency slot set size *e* is constructed based on Whiteman's generalized cyclotomy. To the best of our knowledge, this is the first paper which uses Whiteman's generalized cyclotomy for the construction of FH sequences. Based on some properties of Whiteman's generalized cyclotomy, the proposed FH sequences' Hamming correlation distribution is determined completely. It is shown that the FH sequence set is optimal with respect to the average Hamming correlation bound. Moreover, the proposed FH sequences have large period such that they can be used as a good candidate for military applications.

## 7  References

[1] Scholtz, R.A.: 'The spread spectrum concept', *IEEE Trans. Commun.*, 1977, COM-25, pp. 748-755

[2] Dixon, R.C.: 'Spread spectrum systems' ( New York: Wiley, 1976)

[3] Lempel, A. and Greenberger, H.: 'Families of sequences with optimal Hamming correlation properties', *IEEE Trans. Inform. Theory*, 1974, 20, (1), pp. 90-94

[4] Peng, D.Y. and Fan, P. Z.: 'Lower bounds on the Hamming auto- and cross-correlations of frequency- hopping sequences', *IEEE Trans. Inform. Theory*, 2004, 50, (9), pp. 2149-2154

[5] Peng, D.Y., Niu, X.H., Tang, X.H., and Chen, Q.C.: 'The average Hamming correlation for the cubic polynomial hopping sequences', International Wireless Communications and Mobile Computing Conference (IWCMC 2008), Crete Island, Greece, Aug 2008, pp. 464-469

[6] Peng, D.Y., Niu, X.H., and Tang, X.H.: 'Average Hamming correlation for the cubic polynomial hopping sequences', *to be published in IET Communications*

[7] Chu, W. and Colbourn, C.J.: 'Optimal frequency-hopping sequences via cyclotomy', *IEEE Trans. Inform. Theory*, 2005, 51, (3), pp. 1139-1141

[8] Ding, C., Moisio, M.J., and Yuan, J.: 'Algebraic constructions of optimal frequency-hopping sequences', *IEEE Trans. Inform. Theory*, 2007, 53, (7), pp. 2606-2610

[9] Ding, C. and Yin, J.: 'Sets of optimal frequency-hopping sequences', *IEEE Trans. Inform. Theory*, 2008, 54, (8), pp. 3741-3745

[10] Ge, G.N, Miao, Y., and Yao, Z.X.: 'Optimal frequency hopping sequences: auto- and cross- correlation properties', *IEEE Trans. Inform. Theory*, 2009, 55, (2), pp. 867-879

[11] Fuji-Hara, R., Miao, Y., and Mishima, M.: 'Optimal frequency hopping sequences: a combinatorial approach', *IEEE Trans. Inform. Theory*, 2004, 50, (10), pp. 2408-2420

[12] Peng, D.Y., Peng, T., and Fan, P.Z.: 'Generalized class of cubic frequency-hopping sequences with large family size', *IEE Proceedings on Communications*, 2005, 152, (6), pp. 897-902

[13] Peng, D.Y., Peng, T., Tang, X.H., and Niu, X.H.: 'A class of optimal frequency hopping sequences based upon the theory of power residues', Sequences and Their Applications (SETA 2008), Lexington, KY, USA, September 14-18, 2008, pp. 188-196

[14] Whiteman, A.L.: 'A family of difference sets', Illinois J. Math, 1962, 6, pp. 107-121

[15] Han, Y.K. and Yang, K.: 'New near-optimal frequency-hopping sequences of length *pq*', Proceedings of 2008 IEEE Intl. Symp. Inform. Theory (ISIT 2008), Toronto, Canada, July 2008, pp. 2593-2597

[16] Chung, J.H., Han, Y.K., and Yang, K.: 'New classes of optimal frequency-hopping sequences by interleaving techniques', *IEEE Trans. Inform. Theory*, 2009, 55, (12), pp. 5783-5791

[17] Storer, T.: 'Cyclotomy and difference sets' (Chicago, IL: Marham, 1967)

[18] Bai, E.J, Fu, X.T., and Xiao, G.Z.: 'On the linear complexity of generalized cyclotomic sequences of order four over $\mathbf{Z}_{pq}$', *IEICE Trans.Fundamentals*, 2005, E88-A, (1), pp. 392-395

[19] Ding, C.: 'Linear complexity of generalized cyclotomic binary sequences of order 2', *Finite Fields and Their Applications*, 1997, 3, pp. 159-174

[20] Ding, C.: 'Autocorrelation values of generalized cyclotomic sequences of order two', *IEEE Trans. Inform. Theory*, 1998, 44, (4), pp. 1699-1702

[21] Cusick, T.W., Ding, C., and Renvall, A.: 'Stream ciphers and number theory' (Amsterdam, The Netherlands: Elsevier/North-Holland, 1998), pp. 83-115

[22] Fan, P.Z., Lee, M.H., and Peng, D.Y.: 'New family of hopping sequences for time/frequency hopping CDMA systems', *IEEE Trans. on Wireless Communications*, 2005, **4**, (6), pp. 2836–2842

[23] Jovancevic, A.V., and Titlebaum, E.L.: 'A new receiving technique for frequency hopping CDMA systems: analysis and application'. IEEE 47th Vehicular Technology Conference, May 1997, Vol.3, pp.2187–2190

[24] Maric, S.V.: 'Frequency hop multiple access codes based upon the theory of cubic congruences', IEEE Trans. Aerosp. Electron. Syst., 1990, 26, (6), pp.1035–1039